\documentclass[sigconf]{acmart}

\usepackage{booktabs} % For formal tables
\usepackage{threeparttable}
\usepackage{amsmath}
\usepackage[linesnumbered,ruled]{algorithm2e}
\usepackage{subfig}
\usepackage{balance}

\DeclareMathOperator*{\argmax}{argmax}

\title{Using Image Fairness Representations in Diversity-Based Re-ranking for Recommendations}

\begin{document}

\author{Chen Karako}
\affiliation{%
  \institution{Shopify Inc.}
  \streetaddress{490 Rue de la Gauchetiere O.}
  \city{Montreal}
  \state{QC}
  \postcode{H2Z 0B2}
}
\email{chen.karako@shopify.com}

\author{Putra Manggala}
\affiliation{%
  \institution{Shopify Inc.}
  \streetaddress{490 Rue de la Gauchetiere O.}
  \city{Montreal}
  \state{QC}
  \postcode{H2Z 0B2}
}
\email{putra.manggala@shopify.com}

% \renewcommand{\shortauthors}{C. Karako and P. Manggala}

\begin{abstract}
The trade-off between relevance and fairness in personalized recommendations has been explored in recent works, with the goal of minimizing learned discrimination towards certain demographics while still producing relevant results.

We present a fairness-aware variation of the Maximal Marginal Relevance (MMR) re-ranking method which uses representations of demographic groups computed using a labeled dataset.  This method is intended to incorporate fairness with respect to these demographic groups.

We perform an experiment on a stock photo dataset and examine the trade-off between relevance and fairness against a well known baseline, MMR, by using human judgment to examine the results of the re-ranking when using different fractions of a labeled dataset, and by performing a quantitative analysis on the ranked results of a set of query images.  We show that our proposed method can incorporate fairness in the ranked results while obtaining higher precision than the baseline, while our case study shows that even a limited amount of labeled data can be used to compute the representations to obtain fairness. This method can be used as a post-processing step for recommender systems and search.
% We show that our proposed method is robust against the amount of labeled data used to compute the representations.  This implies that the method can be practically useful even with a limited amount of labeled data and since this method extends MMR, it can be used as a post-processing step for recommender systems and search.
\end{abstract}



\maketitle

\section{Introduction}

Recommender systems have become prevalent in our daily lives, influencing the movies and music we choose to enjoy, to the news and job postings we see. Since recommended items are traditionally ranked in order of relevance to the user, this can give rise to issues of fairness and bias that may be learned from data. In order to ensure equal opportunities and avoid reinforcing bias in systems, there have recently been various efforts to provide fairness-aware recommendations. 

In this work, we expand on some of these earlier efforts by incorporating image data in order to increase fairness in a sorted set of images. We propose a fairness-aware variation of Maximal Marginal Relevance (MMR) re-ranking to combine fairness and relevance when retrieving untagged images similar to an input image or query. We construct fairness representations using deep net image descriptors on a small sample of curated labeled data, and describe how these can be used in the proposed method to achieve fairness over any property or demographic. Our method can be implemented modularly as a post-processing step to any ranking technique that outputs item relevance, including personalized recommender systems.

Our method allows for a trade-off between relevance and fairness, which allows the system designer to vary the amount of fairness in the results. This decision could be based on the context and amount of personalization in the image retrieval system. Depending on the type of image retrieval system, the designer can choose how much fairness should factor into the results. For example, a system meant for retrieving images of a specific person (e.g. looking for ``all photos of my sister'' on a photo application), would likely not need a fairness term. On the other hand, an image retrieval system meant to help users find stock photos for their website after they upload images of a certain theme (e.g. fitness) or style (e.g. pastel colors), may indeed benefit from including fairness, as different demographic groups are represented in the search results.%Such a system would, for example, provide the user with photos of male and female athletes. %On the other hand, a given system can also learn each user's intent and personalize the amount of fairness.

\section{Related Work}
\label{Related Work}
There exist various pre-trained convolutional neural network (CNN) models, such as Inception-v3 \cite{szegedy2016rethinking} and VGG-16 \cite{simonyan2014very} that can be leveraged to extract vector representations of images. These embeddings capture learned image representations and can be used for computing image similarity and performing image search \cite{johnson2017billion}.  Deep architectures, such as those in \citet{simonyan2014very} and \citet{szegedy2016rethinking}, have been shown to be able to learn hierarchical representations with multiple levels of abstraction \cite{lee2010unsupervised}.  

Various efforts in incorporating fairness into ranking and recommendations have previously been pursued.  In \citet{zehlike2017fa}, the authors defined a group fairness criterion and an efficient standalone algorithm for producing fair ranking.  Similarly, \citet{singh2018fairness} proposed a framework that adds fairness constraints on rankings.  Amongst a few group fairness constraints, they defined a demographic parity constraint, which allows the system to enforce that the average promotion of items in some predefined demographic groups is equal. In \citet{celis2017ranking}, the authors presented fast exact and approximation algorithms and hardness results for a constrained ranking maximization problem where fairness notions are incorporated in their constraints. \citet{asudehy2017designing} proposed a system that assists users in choosing weights that induce fairness into the ranking. 

\citet{burke2017balanced} incorporated fairness directly into SLIM \cite{ning2011slim} by adding an extra regularization term to the loss function, called the \textit{neighborhood balance} term, which requires a definition of a \textit{protected} class of users.  Given such a class, this method builds balanced neighborhoods for both the protected and unprotected users, which allows a collaborative filtering algorithm like SLIM to be fairer to the protected users.  Related to this is the concept of \textit{fairness representation} in \citet{zemel2013learning}, which 
enabled the authors to propose a learning algorithm for classification which achieves fairness with respect to protected classes.

One of the most popular result set diversification methods is Maximal Marginal Relevance (MMR) \cite{carbonell1998use}.  MMR allows its user to find items that are both relevant and diverse. \citet{carbonell1998use} showed that documents re-ranked using MMR covered more diverse topics and were more useful in an information retrieval task.  The differential utility of diverse retrieval via MMR and relevance-only retrieval inspired our work, as we frame fairness as a subconcept of diversity.

\section{Incorporating fairness into MMR using image representations}
\label{Incorporating fairness into MMR using image representations}

\subsection{Maximal Marginal Relevance (MMR) retrieval}
We start by describing in detail the MMR algorithm mentioned in the previous section.  Formally, given an item set $\mathcal{I}$ that can be retrieved, we aim to retrieve an optimal subset of items $S^*_k \subset \mathcal{I}$ (where $| S^*_k | = k$ and $k < \left| \mathcal{I} \right|$) that is both relevant and diverse.  Given an item $i \in \mathcal{I}$, the term $\text{rel}(i)$ measures the relevance of $i$, while the term $f(S^*_{j-1} \cup \{i\}) - f(S^*_{j-1})$ measures the gain in diversity after $i$ is added to $S^*_{j-1}$. A parameter $\lambda\in [0,1]$ is used to create a convex combination of these two quantities.  This tunable parameter affects the importance of relevance and diversity of the results.  The pseudo-code and optimization function that is central to this approach is described in Algorithm~\ref{MMR}.

\begin{algorithm}[hb]
    \SetKwInOut{Input}{Input}
    \SetKwInOut{Output}{Output}
%     \underline{Algorithm MMR} $(\mathcal{I}, k, \lambda)$\;
    
    \Input{item set $\mathcal{I}$, size of the desired result set $k$, and convex combination parameter $\lambda$.}
    \Output{list of recommended items $S^*_k$}
    {
    $S^*_0 \leftarrow \emptyset$
    }
    
    {
	\text{for $j \in 1 \ldots k$ do}
    \begin{align*}    
     i_j & = \argmax_{i\in \mathcal{I}\backslash S^*_{j-1}} \lambda \text{rel}(i) + (1-\lambda)\big(f(S^*_{j-1} \cup \{i\}) - f(S^*_{j-1})\big) \\
    S^*_j & = S^*_{j-1} \cup \{i_j\}
    \end{align*} 
    }
    
    {
	  return $S^*_k$;
    }
    \caption{MMR}
\label{MMR}    
\end{algorithm}

Some implicit approaches assume that items covering similar topics should be penalized.  For example, \citet{yu2014latent} assigns:
\begin{align}
\label{Sim}
f(S^*_{j-1} \cup \{i\}) - f(S^*_{j-1}) = - \max_{i'\in S^*_{j-1}} \text{Sim}(i, i'),
\end{align}
which diversifies using pairwise similarity function $\text{Sim}$.  Items that are too similar to an already chosen item are penalized.

This method has a useful practical implication.  If a user has multiple facets in their profile that should be covered by their top-k recommendations, or is a cold-start user that has not been present long enough in the system, MMR re-ranking as a post-processing step to a recommendation algorithm could cover a variety of the different possible topics of the items in the top-k recommendations, which increases the likelihood of relevance for these users.  For a theoretical perspective on diverse retrieval, refer to \citet{sanner2011diverse}.

Note that rel$(i)$ is a general ranking function that depends on the task at hand, for example, this function could represent relevance in search results based on query images or keywords.

\subsection{Image fairness representations via CNN descriptors}
\label{Image fairness representations via CNN image descriptors}

As described in Section~\ref{Related Work}, pre-trained CNN models can be used to extract vector representations of images. In order to obtain a more expressive representation of each image that is higher-dimensional than the 1000 ImageNet output labels, we transform the original Inception-v3 network \cite{szegedy2016rethinking} by removing the output layer, such that an image inference yields a 2048-dimensional dense vector.

We run this transformed network on images with membership labels from different demographic groups. We create image fairness representations by averaging the obtained image embeddings per demographic group. Each image fairness representation is a vector which summarizes all images within the same demographic group. Note that this is different from the fairness representations defined in \citet{zemel2013learning}.

For example, consider a binary gender demographic grouping, with one group representing men and the other group representing women. We can use the above to produce two fairness representations by averaging the embeddings of images in each group. This procedure requires demographic membership labeling for these images.

\subsection{Fairness Maximal Marginal Relevance (FMMR) retrieval}
\label{Fairness Maximal Marginal Relevance (FMMR) retrieval}

MMR is agnostic to the similarity metrics used, which allows us to leverage its diversification capabilities to incorporate fairness in the ranking/recommendation output.  We do so by redefining the function Sim from Equation~\ref{Sim} in the following way:
\begin{align}
\text{Sim}(k, k') & = \sum_{v \in \mathcal{I}_\text{rep}} - \left| d(k, v) - d(k', v) \right|,
\label{FSim}
\end{align}
where image fairness representations $\mathcal{I}_\text{rep}$ is a set of vectors obtained via the procedure described in Subsection~\ref{Image fairness representations via CNN image descriptors} and $d(k, v)$ computes the Euclidean distance between vectors $k$ and $v$ from the same space as image fairness representations.

In Equation~\ref{FSim}, given a fairness representation $v \in \mathcal{I}_\text{rep}$, the absolute difference of the distance between $k$ and $v$ and the distance between $k'$ and $v$ represents the difference in the similarities between $k$ and $v$ and between $k'$ and $v$, i.e., the difference between $k$ and $k'$'s contributions to fairness for the demographic group represented by $v$.  The function Sim in Equation~\ref{FSim} is the sum of the negative of these absolute differences across all fairness representations.  A larger magnitude of Sim in Equation~\ref{FSim} implies that $k$ and $k'$ would have similar contributions to fairness. This allows the algorithm to ensure that every newly added item is the item that contributes the most to a gain in fairness.

\begin{algorithm}[ht] 
    \SetKwInOut{Input}{Input}
    \SetKwInOut{Output}{Output}
    
    \Input{images item set $\mathcal{I}$, size of the desired result set $k$, trade off parameter $\lambda$, transformed Inception-v3 model inference $F$, and fairness representations $\mathcal{I}_\text{rep}$.}
    \Output{list of recommended items $S^*_k$}
    
    {
    $S^*_0 \leftarrow \emptyset$
    }
    
    {
	\text{for $j \in 1 \ldots k$ do}    
    \begin{align*}    
     i_j = & \argmax_{i\in \mathcal{I}\backslash S^*_{j-1}} \Big( \lambda \text{rel}(i) - (1-\lambda) \\
     & \times \max_{i'\in S^*_{j-1}} \sum_{v \in \mathcal{I}_\text{rep}} - \left| d(F(i'), v) - d(F(i), v)  \right|\  \Big) \\    
	S^*_j & = S^*_{j-1} \cup \{i_j\}
    \end{align*} 
    }
    
    {
	  return $S^*_k$;
    }
    \caption{FMMR}
	\label{FMMR algorithm}
\end{algorithm}

We name this variation of MMR with Equation~\ref{FSim} the Fair Maximal Marginal Relevance (FMMR) algorithm, detailed in Algorithm~\ref{FMMR algorithm}.  Note that in this description, we
use the notation $F(i)$ to denote the vector representation of image $i$ obtained by using the transformed Inception-v3 network as per in Subsection~\ref{Image fairness representations via CNN image descriptors}.

\section{Experiments}

\subsection{Data, fairness, and relevance setup}
\label{Data and relevancy setup}
We perform a study of our proposed methodology on an internal dataset of 3249 images from Burst~\cite{burstdotshopifydotcom}\footnote{Note that the Burst photo inventory is constantly being updated with new images.}, a free stock photo service offered by Shopify. These images vary in content, and include people, landscapes, objects, animals, and more. Every image is associated with multiple human-curated tags (such as man, woman, dog, fitness, coffee). Although the compiled dataset is not available, all individual images and their tags can be freely accessed online.

We use the transformed Inception-v3 pre-trained model described in Subsection~\ref{Image fairness representations via CNN image descriptors} to compute vector representations of all the images.

\begin{figure}[hb] 
\begin{tabular}{cccc}
\subfloat[Query image]{\includegraphics[width=0.22\columnwidth]{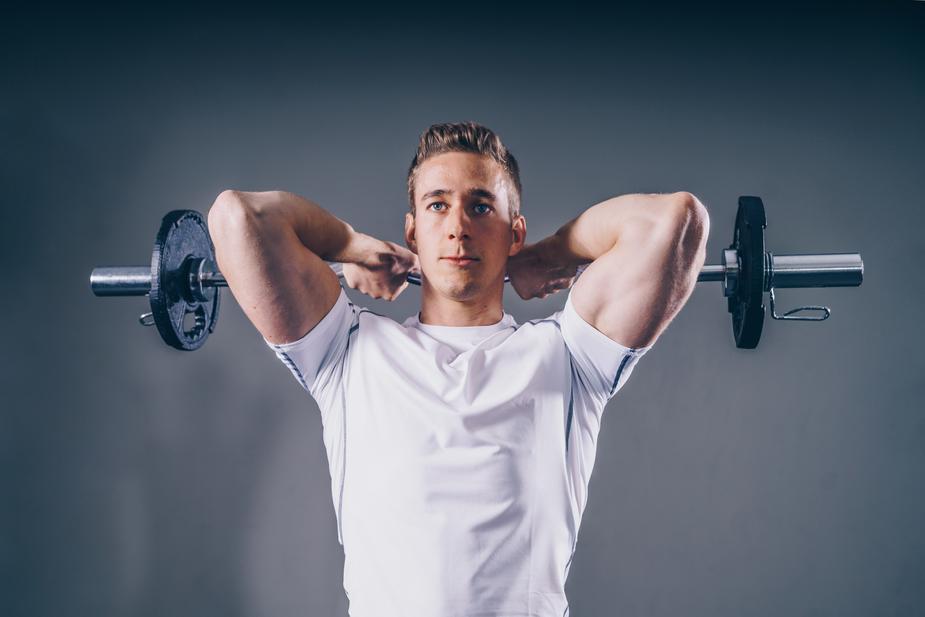}} &
\subfloat[Result 1]{\includegraphics[width=0.22\columnwidth]{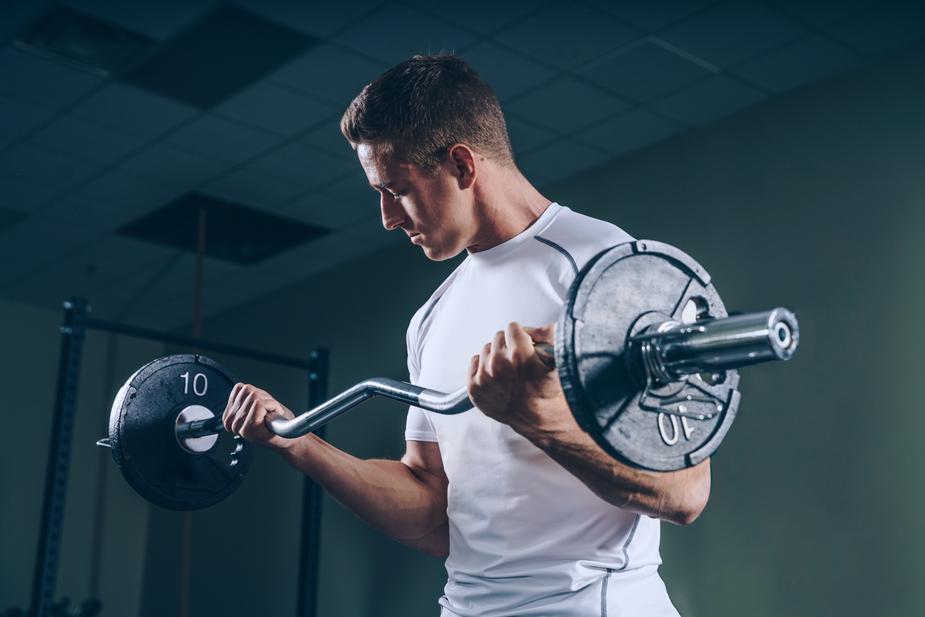}} &
\subfloat[Result 2]{\includegraphics[width=0.22\columnwidth]{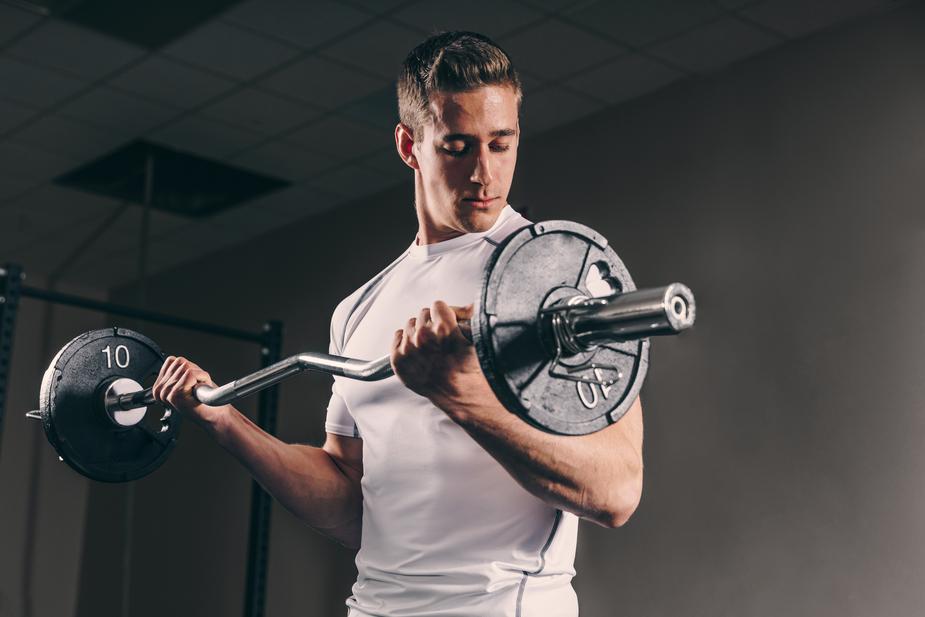}} &
\subfloat[Result 3]{\includegraphics[width=0.22\columnwidth]{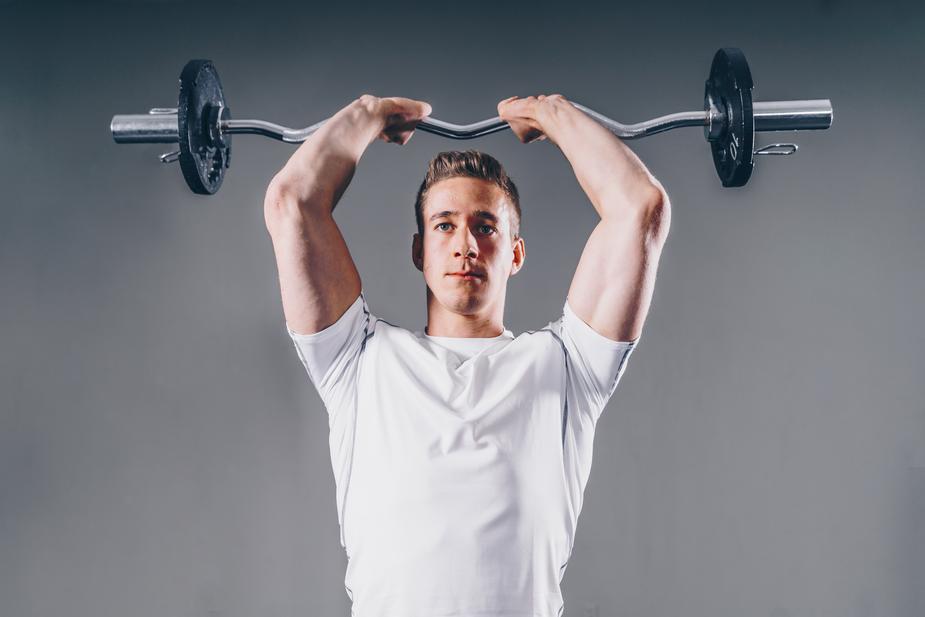}}\\
\subfloat[Result 4]{\includegraphics[width=0.22\columnwidth]{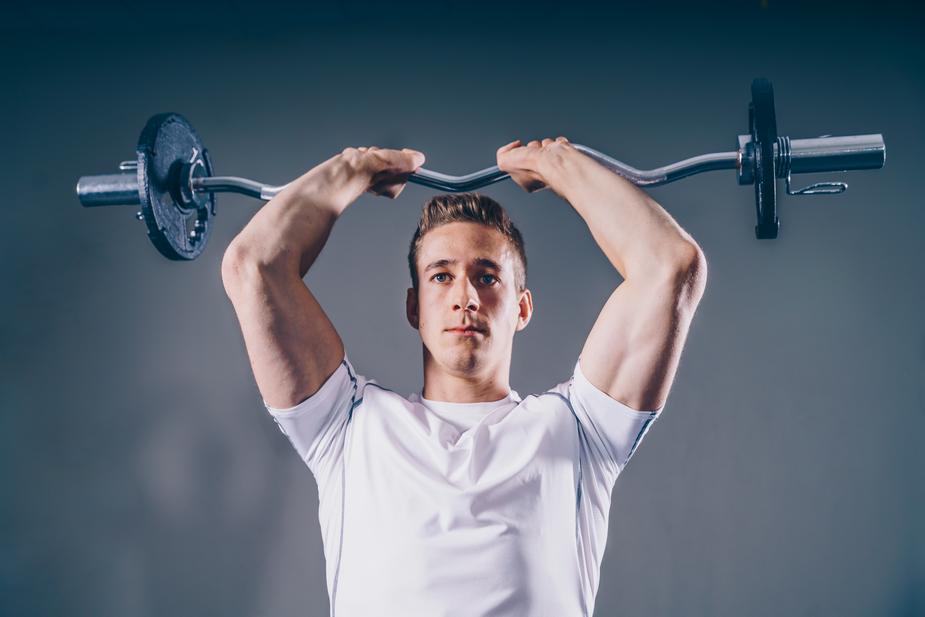}} &
\subfloat[Result 5]{\includegraphics[width=0.22\columnwidth]{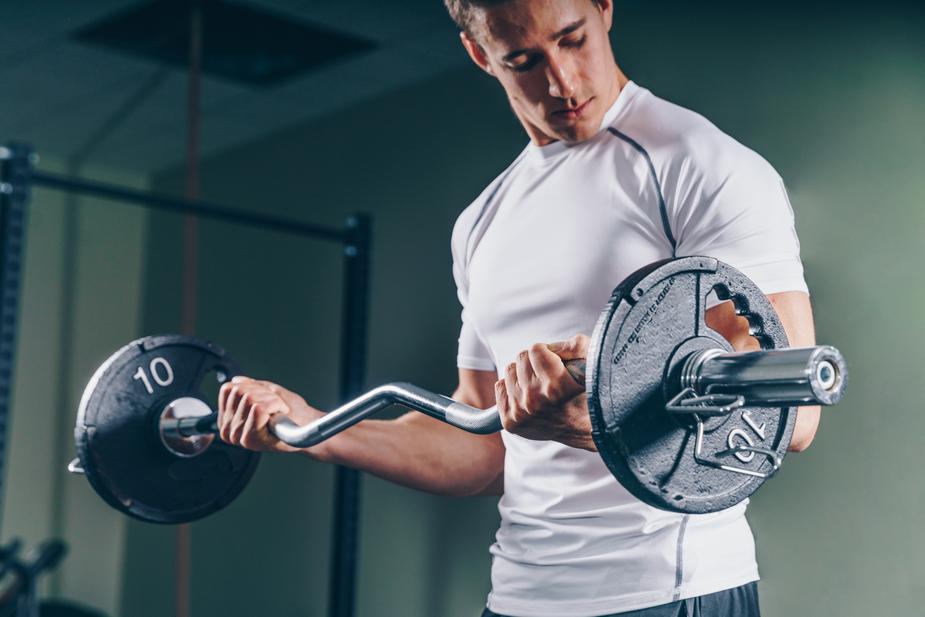}} &
\subfloat[Result 6]{\includegraphics[width=0.22\columnwidth]{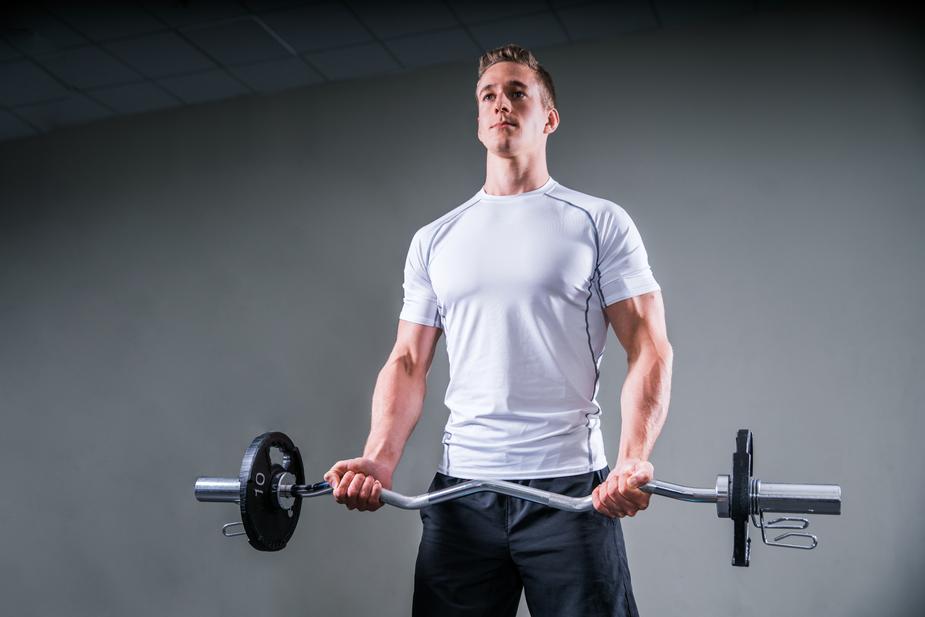}} &
\subfloat[Result 7]{\includegraphics[width=0.22\columnwidth]{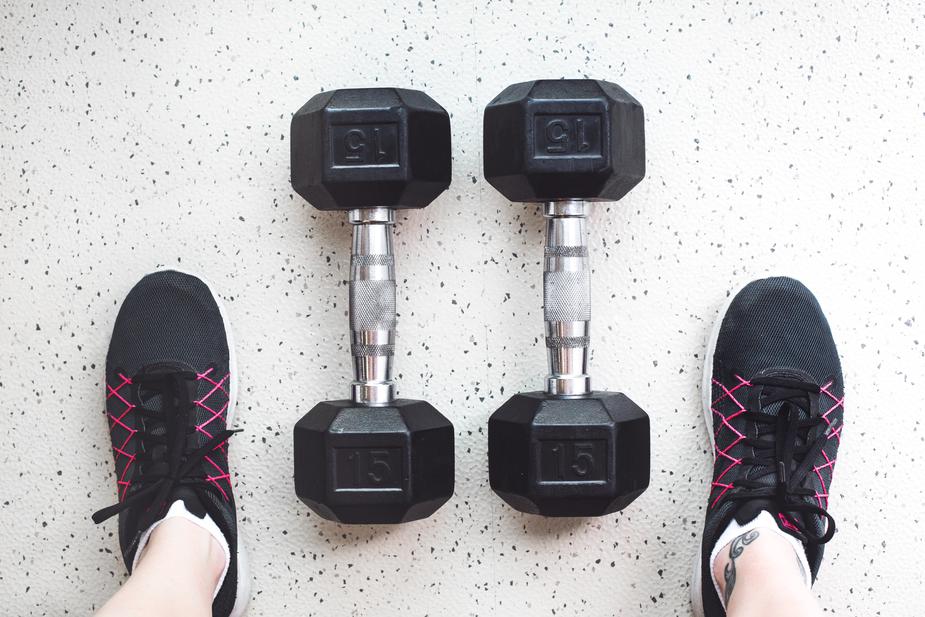}}
\end{tabular}
\caption{Similar images based on KNN ($\lambda=1$). Results are all relevant to the query image but only men are represented, contributing to a fairness ratio at k of 0.}
\label{knn_images}
\end{figure}

The task that we will use for evaluation is similar image retrieval given a query image.  This has been studied in \citet{shankar2017deep} and is very similar to a few commercial large-scale visual search systems in the wild \cite{jing2015visual, yang2017visual}. We choose a query image from the stock photo dataset (for example, the image in the top left of Figure~\ref{knn_images}) and similar to the discussion in \citet{johnson2017billion}, we use k-nearest neighbor (KNN) search using pairwise Euclidean distances between the vector representations of the query image and all the images, to compute the top-k most similar images to the query image, see Figure~\ref{knn_images}. 

We use our proposed algorithm, FMMR, to perform a re-ranking of the top KNN images. Note the setting in Figure~\ref{knn_images} is equivalent to the case where FMMR is applied with $\lambda = 1$. The negative Euclidean distance between the query image and each of the images returned by KNN is used as the relevance function in FMMR.  We have also tried using Manhattan distance, but the results are very similar to when Euclidean distance is used, and are therefore omitted from the discussion.

We choose gender as the demography whose fairness we wish to optimize and obtain fairness representations $\mathcal{I}_\text{rep}$ for `man' and `woman' by averaging the embeddings of all, or a fraction of, images tagged with male- or female-related labels, respectively. In the following section, we refer to this fraction as the \emph{sampling fraction} and study the effects of varying this fraction.  We argue that obtaining good results in terms of relevance and fairness when a relatively small fraction of labeled images is used implies that this method can be useful when not a lot of labeled data is available.  In total, we use 291 images of men and 458 images of women, and fractions of these, to create the fairness representations.

In order to evaluate the performance of FMMR, we compute precision at k ($p$@k) and fairness ratio at k ($fr$@k), which is defined as the ratio between the number of woman images and the sum of the number of man images and woman images in the top-k ranked results. Since we have two demographic groups, the value 0.5 for $fr$@k is optimal. Note that for larger number of demographic groups, one can consider metrics like the entropy of the empirical distribution. Furthermore, we also perform a re-ranking using MMR, and compare the results to those obtained from our algorithm. Similar to KNN, the negative Euclidean distance between the images is used as the similarity function in MMR.

\subsection{Hyperparameter search}
\label{hyperparameter_search}

Hyperparameter search for $\lambda$ is done by performing a grid search that optimizes for fairness while satisficing a certain guarantee of precision.  We introduce a parameter termed the \emph{allowable degradation ratio}, $d$, which we use to choose $\lambda$s for which $p$@k has at most a relative decrease of $d$ from the $p$@k obtained without re-ranking, i.e., $\lambda=1$.  Given these $\lambda$s, we then pick the one that yields the best $fr$@k.  The parameter $d$ serves a practical purpose where the system designers can choose to bound the decrease of $p$@k due to re-ranking.  We use this heuristic to obtain the best $\lambda$ for the different experiments in the next two subsections.

\subsection{Case study evaluation using human judgment on a select image}

We conduct a case study of FMMR and compare it to the MMR baseline by evaluating results for a select query image. In order to perform the FMMR calculation described in Subsection~\ref{Fairness Maximal Marginal Relevance (FMMR) retrieval}, we use the already computed vector representations from Subsection~\ref{Data and relevancy setup}.

\begin{figure}[b] 
\begin{tabular}{cccc}
\subfloat[Query image]{\includegraphics[width=0.22\columnwidth]{weight-lifting-man_925x}} &
\subfloat[Result 1]{\includegraphics[width=0.22\columnwidth]{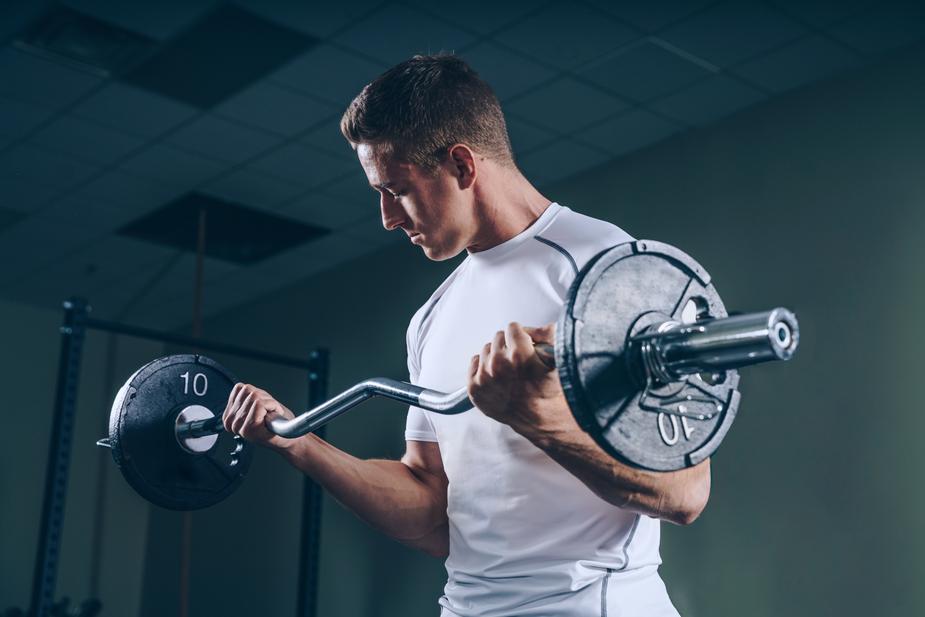}} &
\subfloat[Result 2]{\includegraphics[width=0.22\columnwidth]{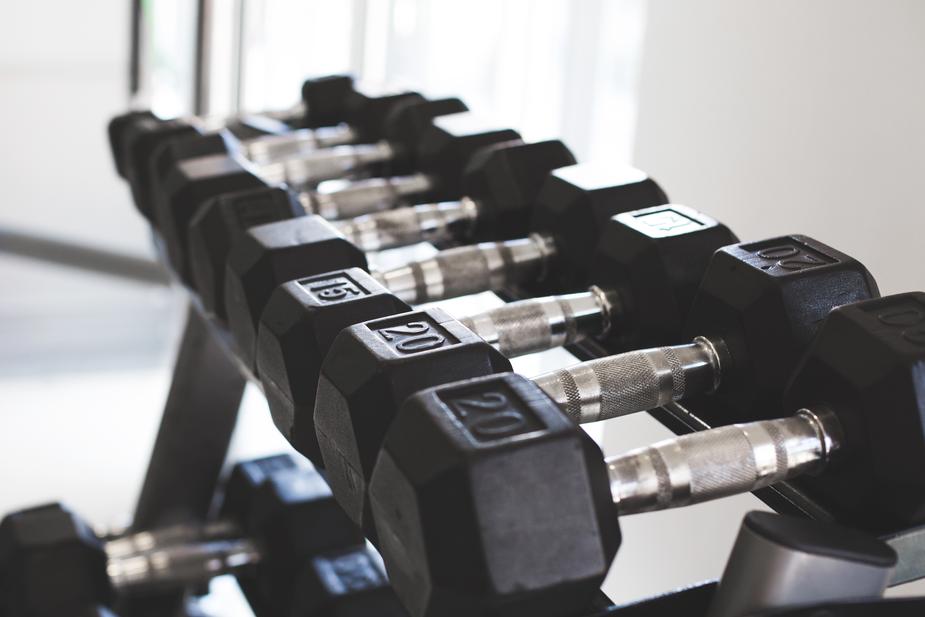}} &
\subfloat[Result 3]{\includegraphics[width=0.22\columnwidth]{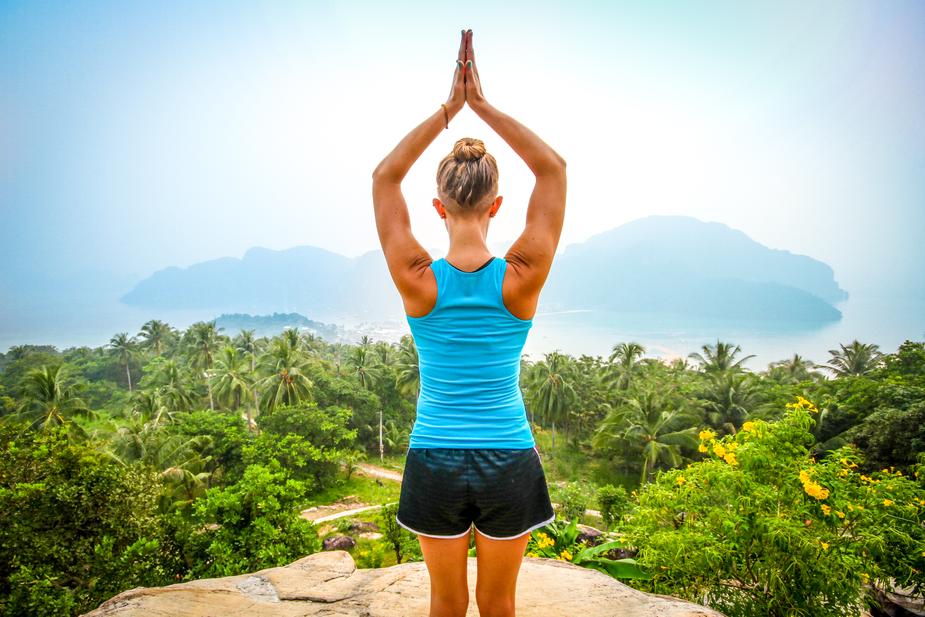}}\\
\subfloat[Result 4]{\includegraphics[width=0.22\columnwidth]{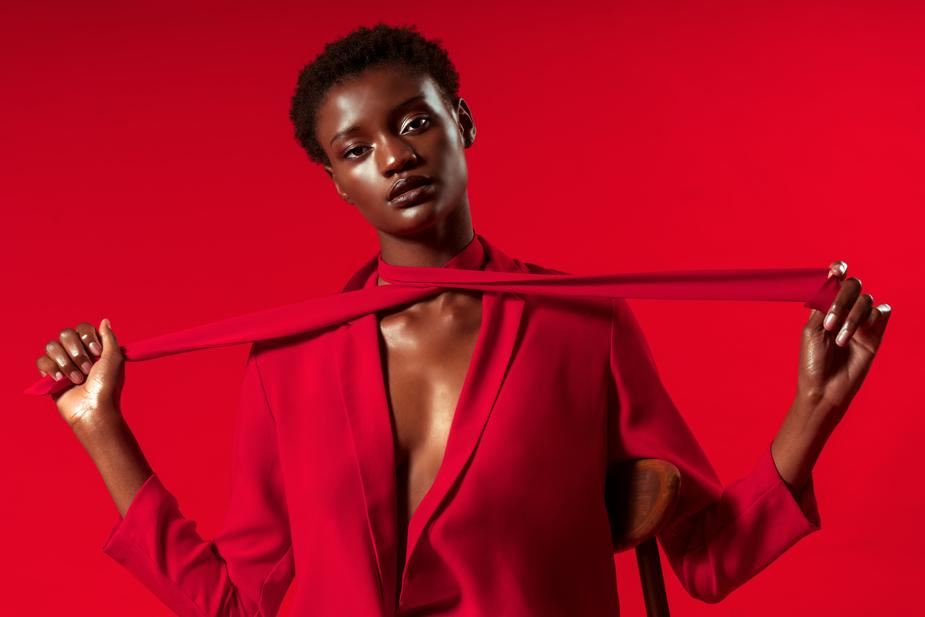}} &
\subfloat[Result 5]{\includegraphics[width=0.22\columnwidth]{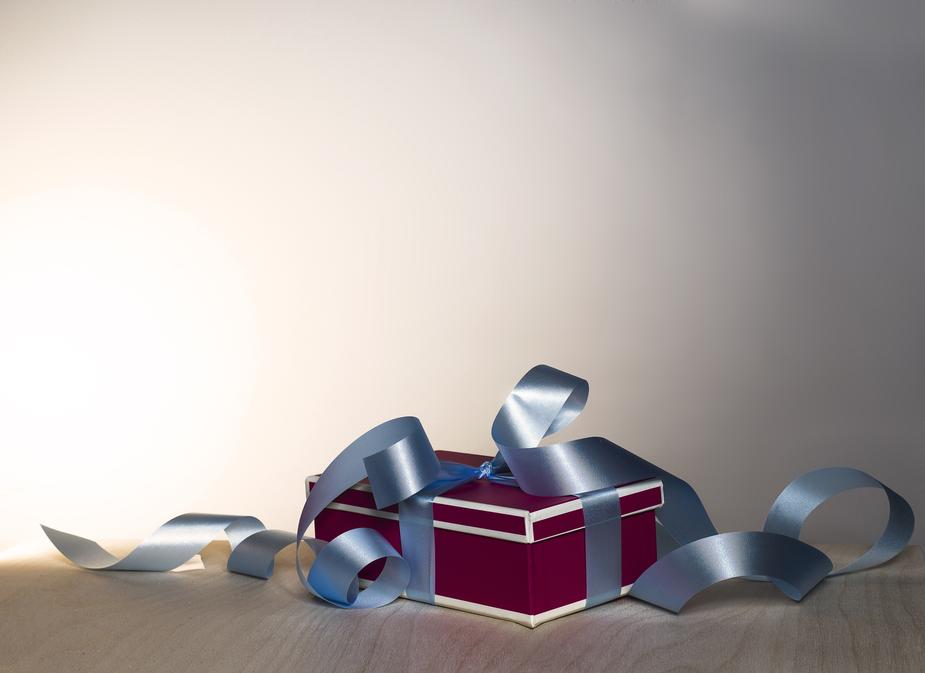}} &
\subfloat[Result 6]{\includegraphics[width=0.22\columnwidth]{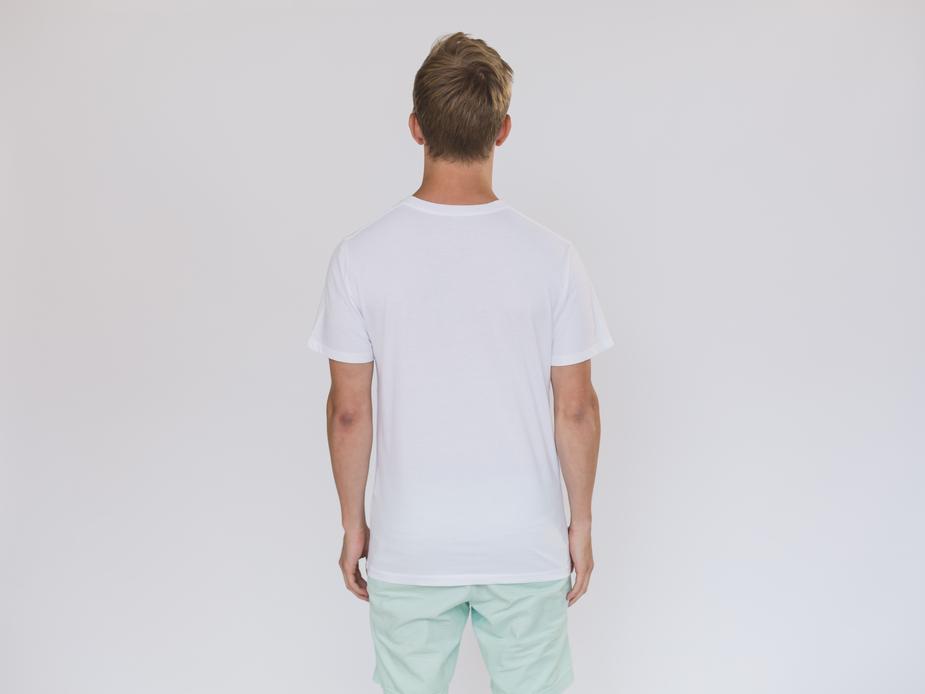}} &
\subfloat[Result 7]{\includegraphics[width=0.22\columnwidth]{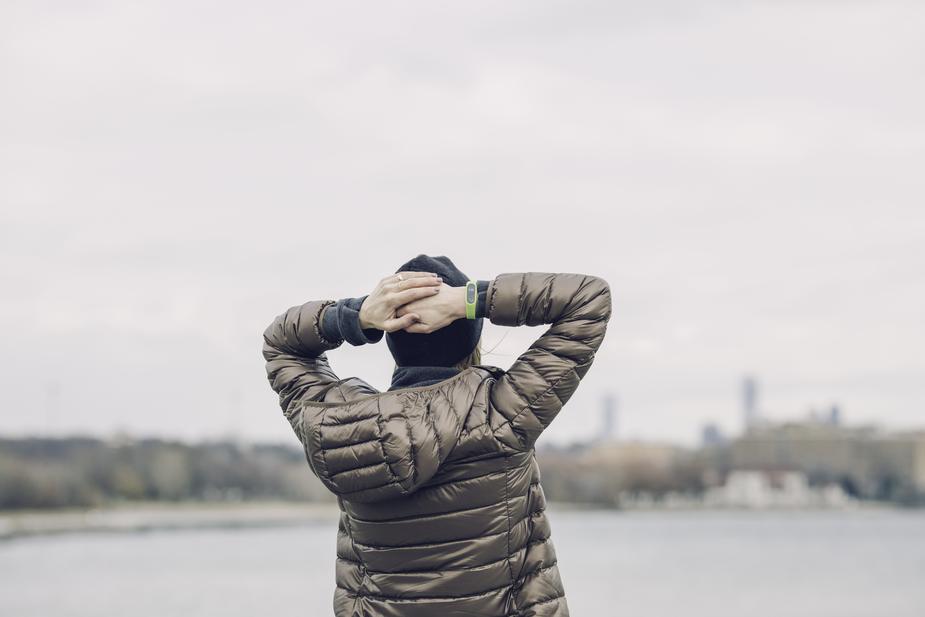}}
\end{tabular}
\caption{Similar images based on MMR ($\lambda=0.40$). Although results show gender diversity, they are not all relevant to the query image.}
\label{mmr_images}
\end{figure}

\begin{figure}[b] 
\begin{tabular}{cccc}
\subfloat[Query image]{\includegraphics[width=0.22\columnwidth]{weight-lifting-man_925x}} &
\subfloat[Result 1]{\includegraphics[width=0.22\columnwidth]{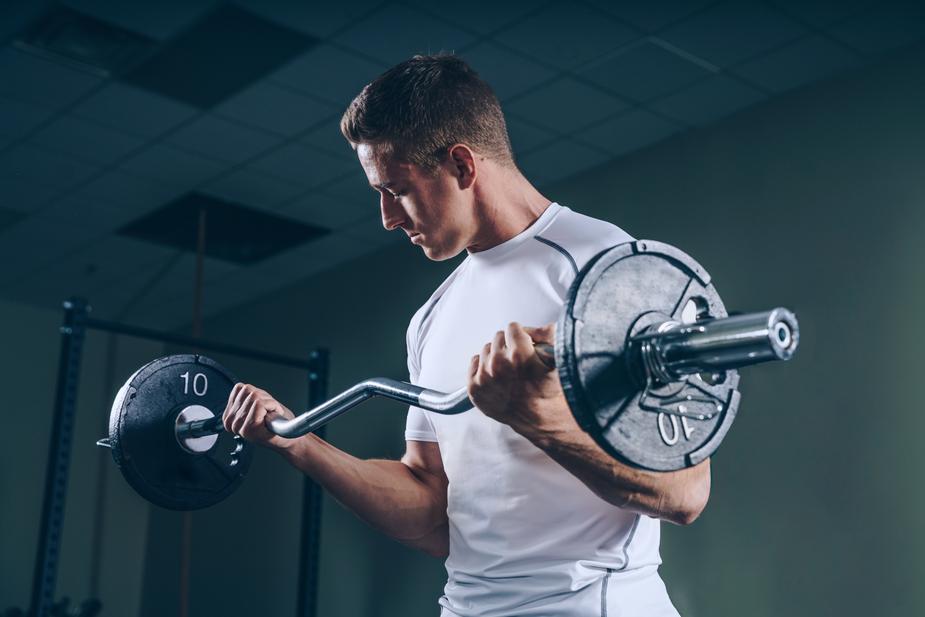}} &
\subfloat[Result 2]{\includegraphics[width=0.22\columnwidth]{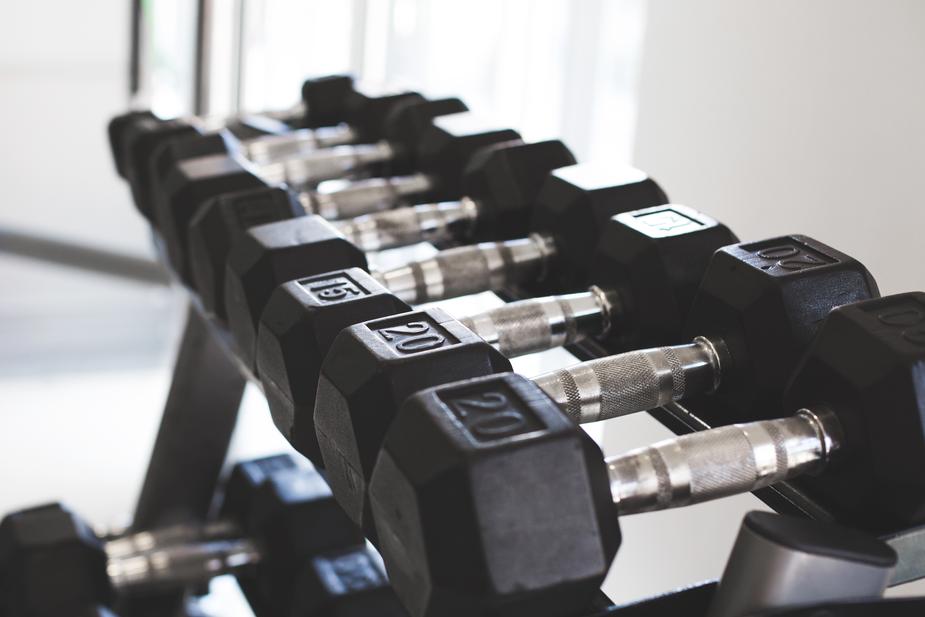}} &
\subfloat[Result 3]{\includegraphics[width=0.22\columnwidth]{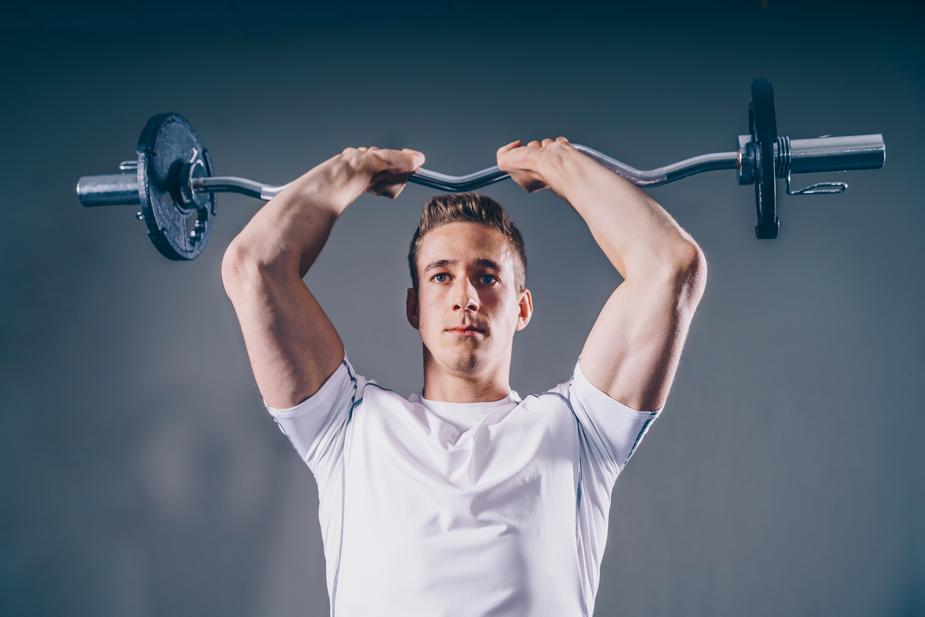}}\\
\subfloat[Result 4]{\includegraphics[width=0.22\columnwidth]{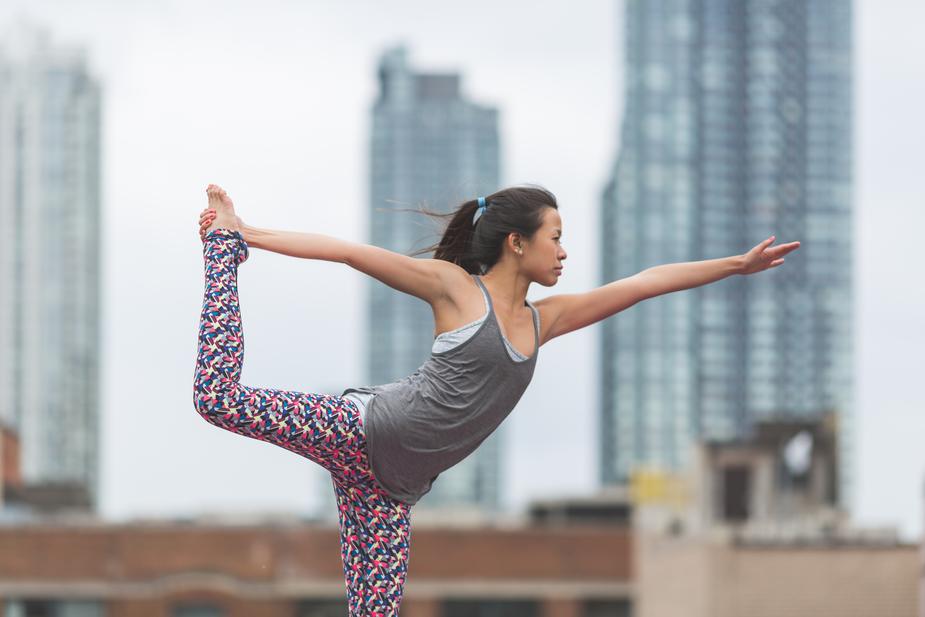}} &
\subfloat[Result 5]{\includegraphics[width=0.22\columnwidth]{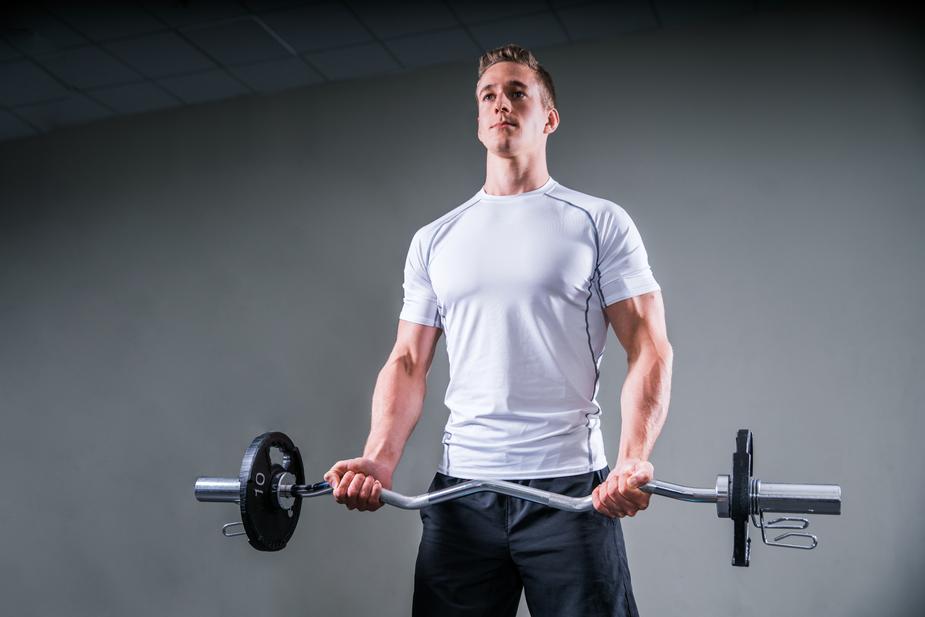}} &
\subfloat[Result 6]{\includegraphics[width=0.22\columnwidth]{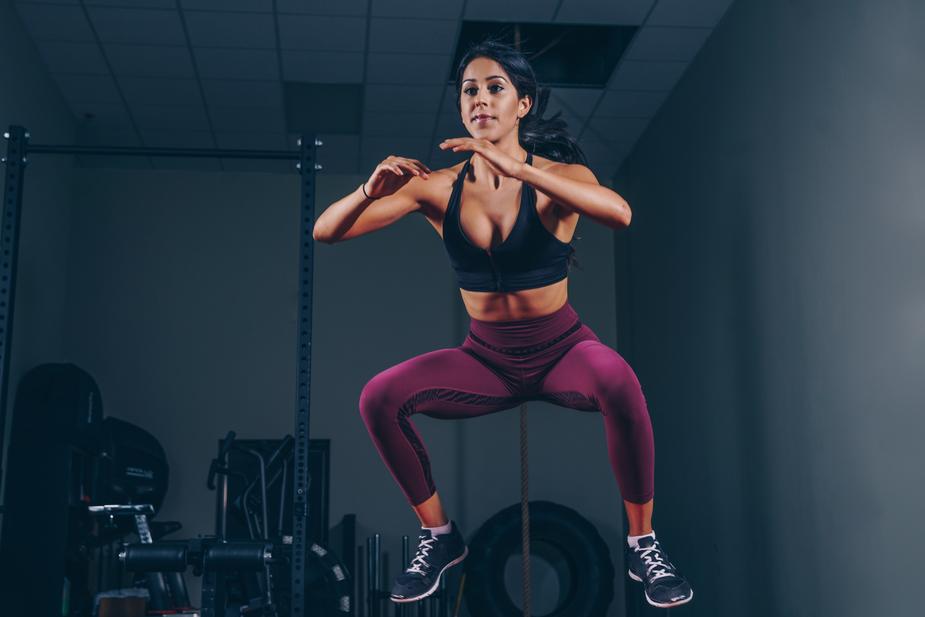}} &
\subfloat[Result 7]{\includegraphics[width=0.22\columnwidth]{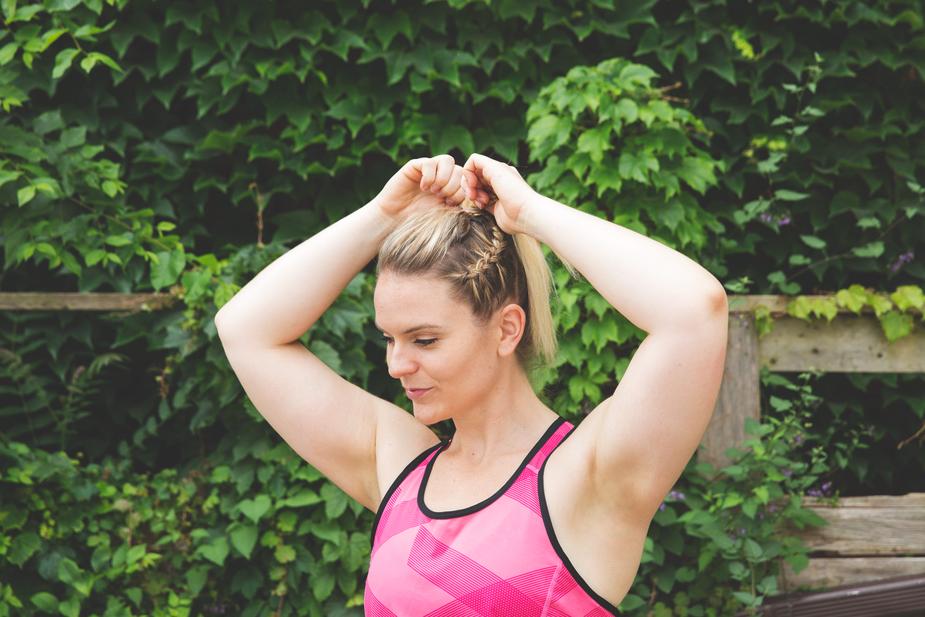}}
\end{tabular}
\caption{Similar images based on FMMR ($\lambda=0.14$, $\mathrm{sampling\ fraction}=1.0$). Results are all relevant to the query image and show gender diversity.}
\label{knn_images_2}
\end{figure}

We use various sampling fractions of labeled data for fairness representation creation, taken from the hyperparameter set $\{ 1, 0.25, 0.1 \}$, where the value 1 means that we use all of the available labeled data. We conduct a hyperparameter search for FMMR, as described in Subsection~\ref{hyperparameter_search}, over $\lambda$ in $[0,1]$, where the value 1 means that we are only considering the KNN relevance, as per in Figure~\ref{knn_images}, and the value 0 means that we are only considering the gain in fairness term.  Using $d=0.25$, which is reasonable for the task at hand, we obtain the optimal $\lambda$ for various sampling fractions for FMMR, and tabulate the precision and fairness of the results. The results of FMMR re-ranking (with $\lambda=0.14$) are shown in Figure~\ref{knn_images_2}.  Qualitatively, we see that these results are still relevant, but now respect the desired demographic parity unlike the KNN-only results in Figure~\ref{knn_images}. The baseline results of MMR re-ranking (with $\lambda=0.40$) are shown in Figure~\ref{mmr_images}.  This value is obtained by searching for $\lambda$ with a similar fairness at k as that obtained from FMMR with the optimal $\lambda$. While the MMR results could achieve demographic parity that is on par with that of FMMR, they are not all relevant to the query.

We assess the quality of the result set by manually examining each image in the ranked list of 10 images in the result set given the query image and qualitatively classifying whether the image is relevant or irrelevant and whether the subject of the image is a man, woman, or neither.  Given these classifications, we can compute a $p@k$ based on human judgment using the former assignment, and $fr$@k using the latter.

\begin{table}[ht]
  \begin{threeparttable}
  \caption{Case study evaluation: $p$@10 and $fr$@10 results}
  \label{tab:results}
  \begin{tabular}{ccccc}
    \toprule
    Re-ranking method & Sampling fraction & $\lambda$    & $p$@10   & $fr$@10 \\
    \midrule
      MMR/FMMR          & N/A               & 1        & 1      & 0              \\ 
      MMR               & 1                 & 0.40     & 0.60   & 0.50           \\ 
      FMMR              & 1                 & 0.14     & 0.90   & 0.50           \\ 
      FMMR              & 0.25              & 0.14     & 0.80   & 0.44           \\ 
      FMMR              & 0.1               & 0.14     & 0.70   & 0.50           \\
      \bottomrule
\end{tabular}
    \begin{tablenotes}
      \small
      \item $p$@10 and $fr$@10 values for different data sampling fractions and $\lambda$s.  The results for the optimal $\lambda$ and sampling fraction for FMMR are shown, along with results for no re-ranking ($\lambda=1$), and baseline MMR ($\lambda=0.40$) for comparison. For each sampling fraction for FMMR, we can obtain reasonably high precision at k while maintaining fairness.
    \end{tablenotes}
\end{threeparttable}
\end{table}

Upon tabulating the qualitative classification results in Table~\ref{tab:results}, we observe that

\begin{itemize}
\item When $\lambda = 1$, FMMR is equivalent to only KNN and all 10 images are relevant ($p$@10 is 1) but the results do not respect demographic parity (fairness ratio at k is 0, since all images are of men). Note that here the concept of sampling fraction is not applicable because we do not make use of the fairness representations. 

% \item For every sampling fraction, as $\lambda$ decreases, the influence of the fairness term in FMMR increases, which decreases the relevance and increases the fairness. 

\item For every sampling fraction, there exists a $\lambda$ that yields precision at k that is at least 0.7 and fairness ratio at k between 0.4 and 0.6, which is only slightly less relevant but significantly improves fairness compared to the case of $\lambda = 1$. 

\item Since we are able to obtain similar values of $p$@10 and $fr$@10 for every sampling fraction, FMMR can be used for various quantities of labeled data, depending on the task at hand; hyperparameter search on $\lambda$ is thus an important step for the application of FMMR.
\end{itemize}

% We note that a thorough quantitative evaluation of our proposed approach was not possible due to incomplete image tags in the dataset.
This case study complements the quantitative evaluation in the next subsection, as we have limited, and sometimes incomplete tags, for the images in the dataset. This thus affects our ability to compute the fairness ratio and precision at k metrics reliably without manual inspection.

\subsection{Quantitative evaluation using labels}

From the full dataset of 3249 images, we choose a set of 738 query images containing image tags that signify the presence of humans in the image, for example, ``men's fashion", ``boys", ``business woman", and ``girls".  We perform hyperparameter search (training) and evaluation on these 738 images instead of the full dataset because the computation of fairness ratio at k relies on the presence of gender-based tags.  Note that images with only non-gender tags are not considered in the fairness ratio calculations. Moreover, each image has multiple tags, and in order to ensure that a search result that is marked as precise is truly relevant in practice, we require at least 1/4 of the available tags of the query image to be matched by a search result. Note that even though the query image is chosen from the set of 738 images, search is performed on all 3249 images in order to reproduce real life setting.

We use a sample of 100 images from this set to perform a hyperparameter search as described in Subsection~\ref{hyperparameter_search}. In the grid search, we check 50 evenly spaced $\lambda$s in the interval $[0,1)$. The best $\lambda$s are obtained from each of the 100 images using the above grid search and are averaged to get one overall best $\lambda$ for each re-ranking method.  For this experiment, we choose $d=0.25$, which is deemed acceptable for the task at hand.  For KNN/relevance function, we choose to only keep the 50 most relevant images for each query image.  Lastly, we consider a result set of size 10 for both hyperparameter search and evaluation.

For each re-ranking method, using the overall best $\lambda$ on 638 test query images that were not used for hyperparameter search, we compute a 95\% t-based confidence interval on the precision at 10 and fairness ratio at 10. The results are tabulated in Table~\ref{tab:label_results}.  Using exactly the same hyperparameter search method and settings, we are able to use FMMR to obtain results that are fair while achieving a significantly higher precision at 10 than MMR.

\begin{table}[ht]
  \begin{threeparttable}
  \caption{Evaluation using labels: $p@10$ and $fr$@10 results}
  \label{tab:label_results}
  \begin{tabular}{cccc}
    \toprule
    Re-ranking method  & $p$@10   & $fr$@10 \\
    \midrule
      MMR            & 0.53 $\pm$ 0.02    & 0.64 $\pm$ 0.02  \\
      FMMR           & 0.59 $\pm$ 0.02    & 0.65 $\pm$ 0.02 \\
  \bottomrule
\end{tabular}
    \begin{tablenotes}
      \small
      \item $p$@10 and $fr$@10 values for MMR and FMMR. The $\lambda$ with best fairness ratio at 10 found for MMR yields a significantly lower precision at 10 than $\lambda$ with the best fairness ratio at 10 found for FMMR. \end{tablenotes}
\end{threeparttable}
\end{table}

\section{Conclusion and Future Work}

We have presented a fairness-aware re-ranking algorithm that allows to define and learn fairness representations which are used to re-rank images. We have shown that this algorithm can produce results with higher precision at k than the MMR baseline while obtaining a good fairness ratio at k.  A case study that we performed has also shown that FMMR is able to obtain fairness despite only learning fairness representations from a limited labeled dataset of tagged images. We note that using the distributed representations instead of classification labels allows us to capture more semantic information useful for computing the distances between a vector and fairness representations. Moreover, this allows us to consider multiple demographics simultaneously as they will all be represented in the same vector space, thus balancing fairness in several facets (such as gender, age, and race).

While we demonstrated our method in an image search task, the generality of the relevance function used in FMMR (and MMR) allows this method to be customized according to the task at hand. For example, this method can be used to rerank results from an image search using a query image or query keyword. 

Since FMMR uses a tunable hyperparameter, the system designers have the option to let users decide how much fairness to incorporate in their search results, or to decide on this trade-off internally in tandem with user studies which seek to discover the general user intention in the platform.

We intend to extend this study by using multiple demographics to compute the fairness representations, running larger scale user studies and utilizing user feedback data to re-rank results from personalization algorithms and search engines. As discussed in Section~\ref{Related Work}, the learned hierarchical representation of data in a deep net means that the closer the representations are to the output labels, the more abstract are the features learned.  Since biases in computer vision techniques have been shown to exist \cite{buolamwini2018gender}, another future work avenue will be to investigate whether using lower level representations in FMMR is less biased and provides better relevance and fairness trade-off than higher level representations. It would also be intriguing experiment with representations based on other pre-trained networks and compare the results. We are also interested in exploring whether our method works for fairness representations that are computed from word embedding methods \cite{goldberg2014word2vec} and joint image-text embedding methods \cite{ren2016joint}.  Lastly, a few fairness measures for ranked outputs have been proposed in \citet{yang2016measuring} and we intend to explore the performance of FMMR further using these measures.

\begin{acks}
  The authors would like to thank Peng Yu for his continued collaboration and infrastructure support. The authors would also like to thank the anonymous referees for their valuable comments and helpful suggestions.
\end{acks}

\bibliographystyle{ACM-Reference-Format}
\balance
\bibliography{sample-bibliography}

\end{document}